\documentclass[5p,times]{elsarticle}

\usepackage{subfigure}
\usepackage{amsmath}
\usepackage{amssymb}
\usepackage{tabularx}
\usepackage[table]{xcolor}
\usepackage{adjustbox}
\usepackage{array}
\usepackage{pgfplots}
\usepackage{bchart}
\usepackage{times}
\usepackage{multirow}
\usepackage{cancel}
\usepackage{tikz}
\usepackage{url}
\usetikzlibrary{positioning,calc,decorations.pathreplacing,arrows.meta,bending}

\usepackage{mathtools}

\DeclarePairedDelimiter\floor{\lfloor}{\rfloor}

\usepackage{algorithm}
\usepackage{algpseudocode}
\usepackage{multicol}
\newlength{\adrfcommentwidth}
\newlength{\adrfstatementwidth}
\newcommand{\adrfstatecomment}[2]{%
  \settowidth{\adrfstatementwidth}{#1}%
  \settowidth{\adrfcommentwidth}{$\triangleright$~#2}%
  \ifdim\adrfcommentwidth>\dimexpr\linewidth-\adrfstatementwidth-1em\relax
    \setlength{\adrfcommentwidth}{\dimexpr\linewidth-\adrfstatementwidth-1em\relax}%
  \fi
  \State #1\hfill\parbox[t]{\adrfcommentwidth}{\raggedleft\strut$\triangleright$~#2\strut}%
}

\journal{arxiv.org}

\begin{document}

\begin{frontmatter}



\title{Autonomous Dominant Resource Fairness for Blockchain Ecosystems}


\author[1]{Serdar Metin\corref{cor1}}
\ead{balikakli@gmail.com}

\cortext[cor1]{Corresponding author}

\address{Alçakdam Yokuşu Sokak 9/7 Cihangir Beyoğlu İstanbul}

\begin{abstract}
Blockchain systems have been a part of mainstream academic research, and a hot topic at that. It has spread to almost every subfield in the computer science literature, as well as economics and finance. Especially in a world where digital trust is much sought for, blockchains offer a rich variety of desired properties, such as immutability, public auditing, decentralised record keeping, among others. Not only has it been a research topic of its own, the integration of blockchains into other systems has been proposed as solutions in many areas, ranging from grid computing, cloud and fog computing, to internet of things, self driving vehicles	, and smart cities. In many cases the primary function attributed to blockchains in these contexts is resource management. Although much attention is paid to this topic, the focus is on single resource allocation scenarios. Even the cases where multiple resource types are to be allocated, are treated as single resource type scenarios, and problems are formulated as allocating standardised bundles consisting of a fixed amount of each of them, such as virtual machines. From a global point of view, this leads to resource waste, since some resources are left idle by the tasks which do not need them, and to which they are allocated along with the resources they need. The present study addresses the problem of allocating multiple resource types among tasks with heterogeneous resource demands with a smart contract adaptation of Precomputed Dominant Resource Fairness; an algorithm that approximates Dominant Resource Fairness within analytically established bounds, without iterative task allocation, which makes it preferable in the blockchain context because of the block gas limit. We present the resulting algorithm, Autonomous Dominant Resource Fairness, along with the empirical data collected from the tests run on the algorithm. The results show that Autonomous Dominant Resource Fairness is a gas-cost efficient algorithm, with gas costs per function call growing linearly with the number of resource types and independently of the number of users. Furthermore, the allocations obtained from the smart contract match those obtained from Precomputed Dominant Resource Fairness in all $11,800$ test cases.
\end{abstract}


\hyphenation{adopt-ed}

\end{frontmatter}

\section{Introduction}
\label{introduction}

For a decade and a half now, blockchain systems have been among the cutting edge research topics. The field emerged when an anonymous author, or a group of authors, with the pseudonym Satoshi Nakamoto \citep{nakamoto2008bitcoin} developed and released the first distributed payment system. 

Bitcoin is developed for addressing the decades old question of how to securely maintain a payment system without the mediation and reassurance of a trusted third party. As such, its functionality is limited to being a decentralised digital currency, or what is commonly referred to as \textit{cryptocurrency}.

The project was a success for achieving its goal. Today, Bitcoin is a widely accepted monetary system throughout the world. But maybe what is more important, and certainly what attracted more attention in academic circles, was the mechanism it employs to maintain a trustless (i.e. not needing a trusted third party), distributed recording system: the blockchain.

Although it offers certain functionalities over simple scripts, generic programmability, which demarks the \textit{second generation} blockchain systems, is not supported in the design of Bitcoin. As such, Bitcoin is accepted to be the only \textit{first generation} blockchain system. The second generation blockchain systems emerged with the development of Ethereum \citep{buterin2013ethereum}. As mentioned above, the main improvement that Ethereum offers is generic programmability, by virtue of \textit{smart contracts}. A smart contract is a script that runs on the virtual machine of the blockchain, and can carry out automatised transactions, alter the state of the virtual machine, make calculations and store data, to name a few of the possible functionalities. As such, it adds great versatility to the blockchain systems.

In contrast to the limited functionalities of the Bitcoin scripts, smart contracts are Turing Complete, meaning, they can run any script that can be run on a Turing Machine; with a limitation, though, which is on the total computational cost of the transactions that can be executed in the processing of a single block.

The mechanism that actualises this limitation is called the \textit{block gas limit}, and it is the main bottleneck in developing blockchain smart contracts. The block gas limit bounds the computational workload associated with a single block, and as such, renders loops with large or unbounded numbers of iterations problematic to implement. It is a predetermined amount, the decision of the value and the update mechanisms of which differs among blockchains.

If a transaction exhausts its available gas during execution, the state changes made during its execution are reverted. Thus, a function whose execution requires more gas than the block gas limit cannot be successfully executed within a single transaction.

Other than the block gas limit, smart contracts are quite permissive to enable a wide range of applications, that are now available in the literature. The functionalities these smart contracts provide allowed blockchains to implement various services, such as decentralised finance (DeFi) \citep{eikmanns2025decentralised}, decentralised autonomous organisations (DAO) \citep{davidson2025nature}, and e-voting \citep{rahul2025articulation}, to name a few, which are under intense research and development.

In addition to the standalone services, blockchains have also been integrated into other systems to handle critical tasks. A common application of blockchains embedded in other systems is resource management, for example in computation clouds and fogs \citep{doka2019cloudagora}, or edge computing and IoT devices \citep{he2020blockchain}. The integration of blockchain systems into these systems provides them with an extra layer of security, privacy, and fairness, which makes them more appealing to the clients they need to attract.

Although resource allocation problems in cloud, fog, and edge computing are addressed with blockchain systems, the main model commonly used assumes a fixed resource bundle, i.e. a standard virtual machine, as the unit of allocation \citep{baranwal2023blockchain}. The present study addresses the same question for the allocation of \textit{multiple resource types}, while keeping fairness both within and between them. The allocation scheme we adopt for this end is Dominant Resource Fairness (DRF) \citep{ghodsi2011dominant}, which is a well established solution in the literature, addressing this scenario.

The main challenge we face at this point is that DRF employs an allocation loop, which is, as indicated before, prone to block gas limit exhaustion problem. We overcome this problem by integrating two mechanisms we developed in prior studies. The first is Autonomous Max-min Fairness (AMF), which implements Max-min Fairness (MF) allocation scheme for single resource types in the blockchain context, by replacing the MF allocation loop with \textit{demand} and \textit{claim} functions executed by the users in a distributed manner \citep{metin2022}. The second is Precomputed Dominant Resource Fairness (PDRF) \citep{metin2025precomputed}, which approximates DRF allocation by precomputing the DRF loop. The algorithm has a computational complexity per function call that is linear in the number of resource types and independent of the number of users. We call the resulting algorithm Autonomous Dominant Resource Fairness (ADRF).

\section{Related Work}
\label{related}

In recent years the blockchain research has spread over many areas such as smart cities \citep{rejeb2021blockchain}, green energy \citep{apeh2025enhancing}, supply chain management \citep{kumar2025blockchain}, e-voting \citep{rahul2025articulation,jayakumari2024voting}, management of medical data \citep{tariq2024revolutionizing}, legislation \citep{wang2024blockchain}, and various applications in banking and finance \cite{gan2025critical}, to name a few. The security and anonymity premises of blockchains offer many advantages in these systems, where social context demands of these highly, such as for personal and commercial privacy.

In addition to founding secure and reliable systems in such a wide range of areas in social systems, blockchains also offer digital trust for high scale information processing systems such as cloud computing \citep{punia2024systematic}, fog computing \citep{solis2024exploring}, edge computing \citep{nguyen2024exploring}, IoT device networks \citep{fazel2024iot}, software defined networks \citep{haritha2024distributed}, and self driving vehicles \citep{tirupati2024blockchain}. Of particular interest to the present study is the handling of fair resource allocation in those systems.

In a survey conducted in 2023, Baranwal and his colleagues \cite{baranwal2023blockchain} reviewed the literature for studies on resource allocation with blockchain based systems. The existing solutions differ from auction based models \cite{doka2019cloudagora, gu2018cloud} to cloud federations \cite{taghavi2018cloudchain}, from negotiation frameworks \cite{pittl2018bazaar} to credit-based management systems \cite{pan2018edgechain}, with algorithms ranging from deep reinforcement learning \cite{feng2019cooperative} to ant colony optimisation \cite{baniata2021pf}. While majority of them run on smart contracts, there are also solutions that offer dedicated proof systems \cite{wilczynski2020modelling}, or even dedicated coins \cite{dou2020blockchain}.

Although there are various models tackling the question of resource management, the studies reviewed in the survey use the abstraction of virtual machines with fixed amounts of different resources.

As it had historically been the case, not much attention is paid to the allocation of multiple resource types, among users with heterogeneous resource demands, while keeping allocation fairness \textit{both within and without}.

Dominant Resource Fairness (DRF) \cite{ghodsi2011dominant} allocation is such a solution, and the present study adapts it to the blockchain context. Prior to the introduction of DRF, resource allocation is mainly done for single resource types. In case of multiple resources, bundles of them were allocated as a single resource, like in blockchain literature, as mentioned above.

DRF is based on Max-min Fairness, which is a widely used fair allocation scheme \cite{hahne1991round, nace2008max, waldspurger1995lottery, gogulan2012max, marbach2002priority}. As such, it also attracted much attention, and is widely adopted in studies such as \cite{li2015note, kash2014no, parkes2015beyond, psomas2013beyond}. It also has been challenged \cite{dolev2012no}, and critically evaluated \cite{gutman2012fair} by studies in the literature.

In a recent study \cite{metin2025precomputed}, we developed Precomputed Dominant Resource Fairness (PDRF), which approximates the DRF allocation within analytically established bounds, and works notably faster than the original algorithm. In the present study we adapt it to the blockchain context\footnote{This may be a bit misleading, since the development happened the other way around. We were trying to adapt DRF to the blockchain context, which lead to an alternative algorithm and the study at \cite{metin2025precomputed} was produced as a consequence. Notwithstanding, it had been completed first, and it is the basis of this study, thus the statement.}.

\section{Reference Models}
\label{model}

In this section we will briefly present the models that the present model was built on. According to that, we will start with the most basic model of Max-min Fairness, in Section \ref{mf}. Based on it we will describe Dominant Resource Fairness in Section \ref{drf}, and in turn, Precomputed Dominant Resource Fairness (\ref{pdrf}). In the following Section (\ref{amf}), we will describe Autonomous Max-min Fairness to introduce a method for blockchain adaptation.

\subsection{Max-min Fairness}
\label{mf}

The defining feature of Max-min Fairness (MF) is maximising the minimum share allocated to any task. The main method for actualising MF scheme is Progressive Filling (PF) algorithm. The working principle of PF is formulised as satisfying the needs of a set of tasks at the same pace; or alternatively formulated as growing each of their utility in equal rates.

For reaching an MF allocation, each task is initially reserved an equal share of the resource. PF starts with the lowest volume demand, and proceeding in the ascending order, allocates each task the minimum of its reserved share and demand, i.e. $\min\left\lbrace d_i, r/n\right\rbrace $ for demand of task $i$, ($d_i$), and the resource reserve $r$, in a demand set of $n$ tasks.

At the end of iteration, the tasks whose demands are lower than or equal to their reserved shares are fully satisfied, and thusly removed from the demand set. The difference between the demands of the removed tasks and their reserved shares are left as residue. The algorithm takes another iteration, reallocating the residual resources $r'$, among the set of remaining tasks in the same manner. Recursively iterating, the algorithm terminates when either all of the tasks are satisfied, or the resource is fully depleted.

The generalised version of MF is Weighted Max-min Fairness (WMF), in which case the tasks are assigned weights depending on some predefined policy, and the resources are reserved for each task proportional to their weight, rather than equally, throughout the iterations. According to this, each task is allocated the minimum of its demand, $d_i$, and its weighted share $ws_i$, which is given by:

\[
ws_i = \frac{w_i}{\sum_j w_j} \cdot r
\]

\noindent for task $i$, and $j \in [1, n]$. In the special case where each task is assigned the same weight, WMF reduces to MF.

\subsection{Dominant Resource Fairness}
\label{drf}

Dominant Resource Fairness (DRF) \cite{ghodsi2011dominant} aims to generalise the main objective of MF to multiple resource scenarios. It does so by introducing the concept of \textit{dominant shares}, and applying the MF principle to maximise the minimum dominant share, $ds$, allocated to any user. The $ds$ of each user $i$ is defined as the highest ratio of her demand for resource $r$, ($d_{ir}$), to the total reserve of the resource:

\[
ds_i = \max_{r \in R} \left\lbrace \frac{d_{ir}}{R_r} \right\rbrace 
\]

\noindent for the resource set $R$, where $R_r$ denotes the reserve of resource $r$. The resource type of the dominant share is defined as the \textit{dominant resource}, $dr$, of the user. The other resources are allocated in their fixed ratio to the $dr$, which is referred to as the \textit{Leontief Preferences}, in the economy literature.

Unlike the MF scenario, in which the users submit their maximum demand for the given resource, the problem is formulated in DRF scenario for the demand of a unit task, and the maximum demand is left open ended. For example, in MF scenario, if a user can complete a task with $2$ units of the given resource, and needs to complete $5$ tasks, she is expected to submit a demand for $10$ units, whereas in DRF scenario, user is expected to submit a demand for $2$ units, and the algorithm allocates as much as it can\footnote{Obviously the DRF scenario can be converted to MF scenario simply by explicitly getting a maximum value from each user. But it should be noted that it is not all that straightforward to do the conversion in the reverse direction. MF scenario cannot be converted simply by getting a unit task from each user, since the indivisibility of tasks brings about new constraints to the problem.}.

Similar to MF, DRF follows the PF principle of growing each user's utility at the same pace. The algorithm iteratively allocates resources to users, picking the demand of the user with the least allocated $ds$, allocating $1$ unit of her task, and putting it back to the demand set with its updated $ds$ allocation. Proceeding in this manner, the algorithm allocates resources until the next selected task cannot be accommodated. Note that a user can be allocated several times consecutively, depending on the ratio between the volume of her dominant share and the volumes of other dominant shares.

With the introduction of a weight vector for each user, DRF can be generalised to its weighted counterpart. The weight vector for user $i$ consists of weights for each resource, $w_i = \left\langle w_{i,1}, w_{i,2}, w_{i,3}, \dots, w_{i,m} \right\rangle $ for each of the $m$ resources, according to some weighting policy. After that it suffices to redefine the $ds$ as the maximum of user demands, divided by the resource reserve \textit{and} its associated weight:

\[
ds_i = \max_{r \in R} \left\lbrace \frac{d_{ir}}{w_{ir} \cdot R_r} \right\rbrace 
\]

\noindent The rest of the algorithm does not need alteration to incorporate the weighting capability.

\subsection{Precomputed Dominant Resource Fairness}
\label{pdrf}

Precomputed Dominant Resource Fairness (PDRF)\citep{metin2025precomputed} operates on the premise of precomputing how many allocations each user would get as a result of DRF iterations, and then assigning that number of tasks to each user at once, without going through the tedious iteration process.

In order to find the number of allocations for each user, PDRF checks the ratios of dominant shares, considering the fact that, each user will get allocation turns inversely proportional to their dominant shares, according to the PF principle. In the same logic, the highest $ds$, $ds^*$, gets the least number of allocations, and in an idealised scenario, where all $ds$'s are integer fractions of $ds^*$, the rest of the $ds$'s gets number of allocations equal to their proportion with respect to it, i.e. each takes $ds^* / ds_i$ allocations, for each allocation $ds^*$ takes. This cycle repeats, until one of the resources is depleted.

Departing from this observation PDRF calculates the amount of reserve drained from each resource in one cycle, and simply by dividing each resource by their relevant rate of depletion, takes the minimum of the results to end up with the number of cycles, $k$, that the algorithm can take before depleting a resource. That value is given by the formula:

\[
k = \min_{r \in R} \left\lbrace \frac{R_r}{\sum_i \frac{ds^*}{ds_i} \cdot d_{ir}} \right\rbrace 
\]

\noindent Each task, in turn, is allocated their demands scaled by, $k$ times the ratio of $ds^*$ to their dominant share, $ds_i$, rounded down:

\[
u_i = \floor*{k \cdot \frac{ds^*}{ds_i}} \cdot d_i
\]

\noindent where $u_i$ denotes the total allocation of user $i$, and $d_i$ denotes her demand vector.

Although in real life scenarios, the unrealistic assumption of all $ds$ values being integer fractions of $ds^*$ rarely holds, if ever, rounding the scaling factor down gives a close approximation. As shown in \citep{metin2025precomputed}, the discrepancy between PDRF and DRF allocations is bounded by
\begin{equation}
	-1 \leq p_i-g_i < \frac{ds^*}{ds_i},
\end{equation}
where $p_i$ and $g_i$ denote the number of tasks allocated to user $i$ by PDRF and DRF, respectively. Accordingly, PDRF underallocates by at most one task, whereas the upper bound on overallocation depends on the ratio of the highest dominant share to the dominant share of the user.

\subsection{Autonomous Max-min Fairness}
\label{amf}

Autonomous Max-min Fairness (AMF) is designed to overcome the block gas limit bottleneck. It is also intended as an exemplary model for replacing centrally executed loops with client side executed functions to distribute the burden of them equally over the users in the blockchain ecosystem.

The operating principle of AMF is not very different from PF. In abstract terms, the overall algorithm still implements an allocation loop, but it is not operated centrally. Instead, the allocation is done by each user for themselves, in epochs, divided into rounds.

Each round emulates the function of one iteration of the main loop. At the beginning of the round, the reserved shares are calculated by the first arriving user\footnote{The gas cost of this operation is refunded to the user for maintaining fairness.}, and each user assigns the minimum of this share and his demand, and deduces the assigned amount from the total reserve, by executing a \textit{claim} function. If the demand is smaller than the reserved share, the user removes herself from the set of demands.

There are two limitations to the algorithm. First, the number of rounds is predetermined. If the reserves are not exhausted at the end of the last round, they are handed over to the next run of allocations. Second, if there are not enough resources to complete a full round (i.e. $r' / n' < 1$), the algorithm should use another policy (e.g. first come first served), or hand the remaining resources over, again, to the next run of allocations.

\section{Autonomous Dominant Resource Fairness Model}
\label{adrf}

Having introduced the main building blocks, we now present Autonomous Dominant Resource Fairness (ADRF). The code of the smart contract implementing ADRF can be accessed at the Github repository \cite{metin2020faucet}.

It should be specified at the onset that ADRF is an implementation of pure PDRF, in which the problem of distributing the excess reserves remaining after the PDRF allocation, is not addressed. The excess reserves are simply handed over to the following executions of the algorithm to be allocated alongside the replenished reserves.

Let us now start by introducing some relevant blockchain notions, continue with giving an overall outline of the algorithm, and then proceed to describing its constituent functions in detail.

\subsection{Timing and Synchronisation}
\label{timing}

In blockchain ecosystems, the time concept is radically different. Unlike the conventional computer systems, which model the solar day for time measurement, blockchains use a discrete time conception, in which each tick corresponds to the inclusion of a new block to the chain. Moreover, the intervals are not rational but ordinal, i.e. unlike the temporal distance of any two solar time units to each other, which is fixed, the temporal distance of inclusion of two blocks may vary wildly. The only temporal measure between two blocks is succession - precedence relation, as defined for another case in \citep{lamport1978time}.

There are two immediate outcomes of this difference:

First, the periods of allocation should be defined in terms of number of blocks in order to synchronise users for carrying out distributed components of the algorithm, as described in Section \ref{amf}, and will similarly be described here.

Second, the resource allocation is not in real time, but it is a token to be used at the convenience of the user it is allocated to, in return for the resource. This also implies that they can be saved and \textit{accumulated} to be used at a later time in the future. For this reason, the handing over of the resources to be allocated in a future time is not an inconvenience particularly in the blockchain context.

\subsection{Fixed-point Arithmetic}

Another constraint of the context is the unavailability of floating point variables in the Solidity programming language. In order to overcome this problem, we employ fixed-point arithmetic by scaling the relevant quantities by a precision variable $p$. In the present study, we set $p$ to $10^6$, corresponding to six decimal places of precision. Integer divisions in the intermediate calculations introduce truncation errors, whose magnitude can be controlled by the choice of $p$. The analytical bounds in Section \ref{pdrf} apply to PDRF in exact arithmetic, rather than to its fixed-point implementation. The implementation assumes positive demand components, nonzero scaled reciprocals and arithmetic without integer overflow. The application of this method will be seen in Sections \ref{update}, \ref{demand}, and \ref{claim}.

\subsection{ADRF outline}
\label{outline}

Operating in such a setting, ADRF consists of three functions, two of which are public, and one private. With the public functions, the users submit their demands for each resource, and claim their shares by assigning the reserved amount to their balance. These functions are named \texttt{demand}, and \texttt{claim}, respectively. The third function, \texttt{update state}, is called by these functions at the beginning of their execution, and it is responsible for updating the machine state. 

The allocation state maintained by the update state function consists of three principal variables:

\begin{itemize}
	\item r: A $2 \times m$ cyclic buffer to hold the reserve information of each resource for two parallel allocated resource pools
	\item $k'$ : Scaled number of iterations the PDRF cycle can take
	\item e : The epoch number
\end{itemize}

The first two of these variables are used for providing the public functions with the necessary information to calculate the allocation values. The last one is used in the synchronisation of the system.

According to the needs of the blockchain setting described above, the time is divided into a fixed number of blocks, which are named epochs. Users are expected to submit their demands in one epoch, and claim their reserved shares in the next. Thus, the number of blocks in an epoch should in the minimum be set to allow each user to make one demand and one claim function calls.

The necessity of the two parallel resource pools governed by a cyclic buffer stems from the need to access the resource reserves in time of registering demands. This information is not available within the execution of an epoch, while users continue claiming their reserved shares and update the reserves accordingly.

The total reserve to be allocated within an epoch is the sum of the excess reserve from a previous epoch, and the replenishment quantity. For this reason, the reserves are kept for different parities of the epoch number separately, and the excess reserve is handed over to the second next epoch, instead of the immediate next, to match the parity of the reserve pool. This way, while the claim function drains one pool, the demand function refers to the reserve of the pool with the complementary parity for registering demands. In the next epoch, the window slides, and the functions switch the pools, to operate in parallel again. At deployment, both resource pools are initialised with the epoch resource vector $er$.

\subsection{Update State}
\label{update}

This is the main function that handles three important components of the algorithm. First, it is responsible for updating the epoch. Second, in the case of an epoch update, it replenishes the resource reserves. Third, again, in case of an epoch update, it is responsible for calculating the number of iterations, $k$.

The pseudo code of \texttt{update state} can be seen in Algorithm \ref{udate_state_pseudocode}.

For updating epoch, the function subtracts the \textit{offset}, the block number immediately following the deployment of the contract, from the block number at the execution time of the function. It then divides the resulting number by the predefined number of blocks for the span of one epoch, which is kept at the constant variable $es$, for epoch span. For convenience, the epochs start from $1$, thus $1$ is added to the result:

\[
e = \floor*{\frac{b - o}{es}} + 1
\]

\noindent where $b$ stands for the block number, and $o$ for offset.

If the resulting number is greater than the present value of the variable, epoch is updated (Algorithm \ref{udate_state_pseudocode}, lines $11$-$12$). In this case, a \textit{selector} variable to select the relevant part of the cyclic buffer of $r$ is initiated with the parity of the epoch (line $13$). The resources are replenished for the complementing parity (line $14$), and the number of cycle iterations is calculated and set for the selector parity (lines $15$-$19$). Conveniently, for the value of their own selector variables, the demand function uses the complementing parity, and the claim function uses the same parity with update state.

\begin{algorithm}[!htbp]
	\caption{ADRF: Update State}
	\label{udate_state_pseudocode}
	\begin{algorithmic}[1]
		\item[\hspace*{.5cm}\texttt{updateState()}]
		\adrfstatecomment{$b$}{Block number}
		\adrfstatecomment{$o$}{Offset}
		\adrfstatecomment{$e$}{Epoch number, starts from $1$}
		\adrfstatecomment{$s$}{Selector for cyclic buffers}
		\adrfstatecomment{$re$}{Cyclic buffer of reset epochs}
		\adrfstatecomment{$ds'^*$}{Cyclic buffer of min. scaled reciprocals}
		\adrfstatecomment{$k'$}{Number of available cycles}
		\adrfstatecomment{$R = \left\langle r_1, r_2, \dots, r_m \right\rangle$}{Cyclic buffer of resource vectors}
		\adrfstatecomment{$ER = \left\langle er_1, er_2, \dots, er_m \right\rangle$}{Vector of epoch reserves}
		\adrfstatecomment{$SDS = \left\langle sds_1, sds_2, \dots, sds_m \right\rangle$}{Cyclic buffer of scaled demand sums}
		\item[]
		\If{$e < \lfloor (b-o)/es \rfloor + 1$}
		\State $e \gets \lfloor (b-o)/es \rfloor + 1$
		\State $s \gets e \mod 2$
		\State $r_{1-s} \gets r_{1-s} + er$
		\If{$re_s \neq e-1$ or $re_s = 0$}
		\State $k' \gets 0$
		\Else
		\State $k' \gets \min_{i \in 1 \dots m} \left\lbrace \lfloor (r_{s,i} \cdot ds'^*_s \cdot p) / sds_{s,i} \rfloor \right\rbrace$
		\EndIf
		\EndIf
		\State return
	\end{algorithmic}
\end{algorithm}

Resources are replenished simply by incrementing the resource variable, $r$, by the fixed amount \textit{epoch resource}, $er$, defined in the deployment of the algorithm as a constant variable:

\[
r_{1 - selector} \gets r_{1 - selector} + er
\]

\noindent For ease of review, from this point on we will represent cyclic buffers in equations as simple variables. Their explicit working may be reviewed in their corresponding pseudocodes.

Note that in cases where multiple epochs have elapsed since the last update, the resources are replenished only once. This policy is adopted for simplicity and is not an inherent requirement of ADRF. Alternatively, each resource pool may be replenished cumulatively according to the number of elapsed epochs associated with that pool, or the entire accumulated replenishment may be directed to the resource pool designated for demand collection in the current epoch.

The epoch in which the scaled demand sums were last reset is recorded separately for each pool. If this epoch does not correspond to the immediately preceding epoch, there are no valid demands for the current distribution, and $k'$ is set to zero.

The number of cycle iterations is given by the formula stated in Section \ref{pdrf}. Before integer truncation, the scaled number of cycles is given by:

\[
p \cdot k = \min_r \left\lbrace \frac{R_r \cdot p \cdot p}{\sum_i \frac{ds^* \cdot p}{ds_i} \cdot d_{ir}} \right\rbrace
\]

Note that we can pull $ds^*$ out of the sum, since it is not dependent on $i$. This allows us to collect $p / ds^*$ and $\sum_i p \cdot d_{ir} / ds_i$ separately. The formula becomes:

\[
p \cdot k = \min_r \left\lbrace \frac{\frac{p}{ds^*} \cdot R_r \cdot p}{\sum_i \frac{p \cdot d_{ir}}{ds_i}} \right\rbrace
\]

\noindent Finally, we replace the dominant share variables with their scaled reciprocals:

\[
ds'_i = \floor*{\frac{p}{ds_i}}, \qquad ds'^* = \min_i ds'_i
\]

\noindent which leaves us with the fixed-point approximation of $p \cdot k$:

\[
k' = \min_r \left\lbrace \floor*{\frac{ds'^* \cdot R_r \cdot p}{\sum_i ds'_i \cdot d_{ir}}} \right\rbrace
\]

\noindent The function iterates over the resource vector, calculating $k'$ values for each resource, and stores the minimum $k'$ value to be used in the allocation. Having stored the $k'$, the function returns.

\subsection{Demand}
\label{demand}

Users register their demands by invoking the \texttt{demand} function, with their demand vector as the argument. The pseudocode of \texttt{demand} can be seen in Algorithm \ref{demand_pseudocode}. For brevity, the registration checks performed by both public functions are omitted from their pseudocodes.

Basically, the function handles $4$ tasks.

\begin{itemize}
	\item It registers the demand vector under the entry for the user in a cyclic buffer (line $16$).
	\item It finds the $ds$ of the user and registers it for the user to a cyclic buffer (line $17$).
	\item It scales the demand vector by $p / ds$ and adds it to the scaled demands sum, $sds$ (lines $20$ and $24$).
	\item It checks the $ds$ of the user against $ds^*$, and updates the latter if needed (lines $21$ and $25$).
\end{itemize}

Note that the last two are variables used by the update state function. While updating these, the function checks a variable, which is named $re$, for \textit{reset epoch}, and indicates when the variables are last reset. The reset epochs are stored in a cyclic buffer, separately for each resource pool. Since at each epoch, old values should be overwritten, if the reset epoch of the selected pool is smaller than the present epoch, rather than checking the $ds^*$ or adding the scaled demand vector to the scaled demand sums, the function sets these variables to the user's own values, and updates the reset epoch of the selected pool to the present epoch.

The epoch in which the user registers her demand is also recorded in a cyclic buffer, separately for each resource pool. This allows the function to prevent repeated demand registrations within the same epoch, and the \texttt{claim} function to verify that the demand was registered in the immediately preceding epoch.

For reasons of fixed point arithmetic, we update the formula to store the \textit{reciprocals} of $ds_i$ and $ds^*$, rather than the variables themselves. The updated variable is then given by:

\[
ds'_i = \floor*{\frac{p}{ds_i}} = \min_r \left\lbrace \floor*{\frac{R_r \cdot p}{d_{ir}}} \right\rbrace 
\]

\noindent and the scaled demands sum, $sds$, is iteratively calculated by each call to the \texttt{demand} function, in a distributed manner. More explicitly, each call updates the running sum for each resource $r$, yielding the sum on the right after all demands have been registered:

\[
sds_r \gets sds_r + ds'_i \cdot d_{ir}, \qquad sds_r = \sum_i ds'_i \cdot d_{ir}
\]

\noindent Similarly, while $ds^*$ is the maximum of the dominant shares, the \texttt{demand} function actually collects the minimum of the scaled reciprocals $ds'_i$ as $ds'^*$.

\begin{algorithm}[!htbp]
	\caption{ADRF: Demand}
	\label{demand_pseudocode}
	\begin{algorithmic}[1]
		\item[\hspace{.5cm}\texttt{demand(d)}]
		\adrfstatecomment{$d$}{Input: demand vector}
		\adrfstatecomment{$s$}{Selector for cyclic buffers}
		\adrfstatecomment{$u$}{User id number, $u \in \left\lbrace 1, \dots, n \right\rbrace$}
		\adrfstatecomment{$re$}{Cyclic buffer of reset epochs}
		\adrfstatecomment{$de$}{Cyclic buffer of users' demand epochs}
		\adrfstatecomment{$ds'^*$}{Cyclic buffer of min. scaled reciprocals}
		\adrfstatecomment{$D = \left\langle d_1, d_2, \dots, d_n \right\rangle$}{Cyclic buffer of demand vectors}
		\adrfstatecomment{$DS = \left\langle ds'_1, ds'_2, \dots, ds'_n \right\rangle$}{Cyclic buffer of scaled reciprocals}
		\adrfstatecomment{$SDS = \left\langle sds_1, sds_2, \dots, sds_m \right\rangle$}{Cyclic buffer of scaled demand sums}
		\adrfstatecomment{$R = \left\langle r_1, r_2, \dots, r_m \right\rangle$}{Cyclic buffer of resource vectors}
		\item[]
		\State \texttt{updateState()}
		\State $s \gets (e + 1) \mod 2$
		\If{$de_{s,u} \geq e$}
		\State revert
		\EndIf
		\State $d_{s,u} \gets d$
		\State $ds'_{s,u} \gets \min_i \left\lbrace \lfloor (p \cdot r_{s,i}) / d_i \rfloor \right\rbrace$
		\State $de_{s,u} \gets e$
		\If{$re_s < e$}
		\State $sds_s \gets d \cdot ds'_{s,u}$
		\State $ds'^*_s \gets ds'_{s,u}$
		\State $re_s \gets e$
		\Else
		\State $sds_s \gets sds_s + d \cdot ds'_{s,u}$
		\State $ds'^*_s \gets \min \left\lbrace ds'^*_s, ds'_{s,u} \right\rbrace$
		\EndIf
		\State return
	\end{algorithmic}
\end{algorithm}

\subsection{Claim}
\label{claim}

Having registered demands in the \texttt{demand}, and calculated number of iterations in \texttt{update state} functions, the \texttt{claim} function now is responsible for calculating user's reserved share and assigning it to her account. The pseudocode of \texttt{claim} can be seen in Algorithm \ref{claim_pseudocode}.

Before calculating the user's reserved share, the function verifies that the user has not already claimed her share in the current epoch and that her demand was registered in the immediately preceding epoch.

The function first calculates the number of allocations, by multiplying the ratio of the $ds^*$ to $ds_i$, by $k$ (lines $21$-$22$). To state it more explicitly and in terms of updated variables and precision factors, the ratio is calculated as such:

\[
ratio = \floor*{\frac{ds'_i \cdot p}{ds'^*}}
\]

\noindent The product $ratio \cdot k'$ is divided by $p \cdot p$ and rounded down to obtain the number of allocations, which in turn is multiplied by the user's demand:

\[
share = \floor*{\frac{ratio \cdot k'}{p \cdot p}} \cdot d_i
\approx \floor*{\frac{ds^*}{ds_i} \cdot k} \cdot d_i
\]

The function returns after assigning the share to the user balance, deducting it from the resource reserves, and updating the user's claim epoch to the present epoch.

\begin{algorithm}[!htbp]
	\caption{ADRF: Claim}
	\label{claim_pseudocode}
	\begin{algorithmic}[1]
		\item[\hspace*{.5cm}\texttt{claim()}]
		\adrfstatecomment{$s$}{Selector for cyclic buffers}
		\adrfstatecomment{$u$}{User id number, $u \in \left\lbrace 1, \dots, n \right\rbrace$}
		\adrfstatecomment{$de$}{Cyclic buffer of users' demand epochs}
		\adrfstatecomment{$ce$}{Users' last claim epochs}
		\adrfstatecomment{$ds'^*$}{Cyclic buffer of min. scaled reciprocals}
		\adrfstatecomment{$k'$}{Number of available cycles}
		\adrfstatecomment{$ratio$}{Dominant share ratio}
		\adrfstatecomment{$share$}{Reserved share}
		\adrfstatecomment{$D = \left\langle d_1, d_2, \dots, d_n \right\rangle$}{Cyclic buffer of demand vectors}
		\adrfstatecomment{$DS = \left\langle ds'_1, ds'_2, \dots, ds'_n \right\rangle$}{Cyclic buffer of scaled reciprocals}
		\adrfstatecomment{$B = \left\langle b_1, b_2, \dots, b_n \right\rangle$}{Vector of user balances}
		\adrfstatecomment{$R = \left\langle r_1, r_2, \dots, r_m \right\rangle$}{Cyclic buffer of resource vectors}
		\item[]
		\State \texttt{updateState()}
		\State $s \gets e \mod 2$
		\If{$ce_u \geq e$}
		\State revert
		\EndIf
		\If{$e = 1$ or $de_{s,u} \neq e-1$}
		\State revert
		\EndIf
		\State $ratio \gets \lfloor (ds'_{s,u} \cdot p) / ds'^*_s \rfloor$
		\State $share \gets \lfloor (ratio \cdot k') / (p \cdot p) \rfloor \cdot d_{s,u}$
		\State $b_u \gets b_u + share$
		\State $r_s \gets r_s - share$
		\State $ce_u \gets e$
		\State return
	\end{algorithmic}
\end{algorithm}

\subsection{Weighting}
\label{weighting}

As noted in \cite{metin2025precomputed}, similar to DRF, a weighted version of PDRF can also be implemented with a weight vector $w_i = \left\langle w_{i,1}, w_{i,2}, w_{i,3}, \dots, w_{i,m}\right\rangle$  assigned to each user $i$, for each of the $m$ resources, and then calculating the $ds$ values with the following, instead of the original formula:

\[
ds_i = \max_r \left\lbrace \frac{d_{ir}}{w_{ir} \cdot R_r}\right\rbrace
\]

\noindent where $w_{ir}$ represents the weight of user $i$ for the resource $r$.

ADRF is no different in this aspect than PDRF. It is possible to assign a weight vector to each user and calculate the dominant share, and in turn the reserved shares, accordingly. It can be seen in the results in Section \ref{results} that ADRF is gas cost efficient enough to include one extra division operation in its constituent functions, and there are no other bottlenecks that would cause a problem. But for reasons of simplicity we did not undertake such an effort, and left it out of the scope of the present study.

\section{Method and Testing Environment}
\label{method}

We carried out the tests on GitHub Actions using Python v. 3.10 and Ganache v. 7.9.2, with a block gas limit of $12,000,000$. Brownie v. 1.21.0 \citep{brownie2020} was installed in the testing environment; the experimental scripts interacted with Ganache directly through Web3.py and compiled the contract using py-solc-x. Each function call is inserted in a block as the only transaction. Thus, the order of the function calls are reflected in block sequence, and synchronised accordingly.

The contract is written in Solidity, compiled with Solidity compiler v. 0.5.17, targeting the Istanbul EVM version, with the optimiser enabled and the number of optimisation runs set to $200$. The scripts to run the tests are written in Python. The demands are drawn from a discrete uniform distribution, from the closed interval $[1,10]$, using a fixed random seed ($20261126$) for reproducibility.

For correctness of calculation, we cross-checked the results with a Python implementation of PDRF employing exact rational arithmetic.

For performance, we took the gas cost of functions as the main metric. We investigated the effects of both the number of users and the number of resources on the gas costs. For the former, we ran the tests with $10$, $100$, and $1,000$ users, keeping the number of resources fixed at $10$. For the latter, we ran the tests with $2$, $5$, $10$, $20$, $30$, $40$, $50$, and $100$ resources, keeping the number of users fixed at $10$.

The results reported here are for the corrected implementation. The initial reserve of each pool is equal to its replenishment quantity. The tests consisted of users generating demands and claiming them in the following epoch, for $10$ times, thus extended over $11$ epochs. Since each function call corresponds to $1$ block, the epoch span is always set to $2n$ blocks. In the first epoch, users are registered for $n$ blocks, and they make \texttt{demand} calls for another $n$ blocks, concluding the epoch. The intermediate epochs follow the sequence of $n$ blocks of \texttt{claim} calls followed by $n$ blocks of \texttt{demand} calls. The final epoch contains only claim calls. Incidentally, all the epoch updates take place in the execution of the first \texttt{claim} call.

For this reason, we recorded the gas costs of \texttt{claim} calls that trigger an epoch update separately from those that do not.

The resource replenishment quantity is set to $150$ units per user for each resource in each epoch, i.e. $1,500$ with $10$ users, $15,000$ with $100$ users and $150,000$ with $1,000$ users.

\section{Results}
\label{results}

Across all experimental configurations, the allocations obtained from ADRF matched those obtained from PDRF in all $11,800$ cases, despite the truncations introduced by fixed-point arithmetic.

According to the data collected from the tests, the gas costs scale linearly with respect to the growing number of resources. Although we collected $10$ runs of data, we are presenting the first $3$ of them in Tables \ref{table:demand_averages} and \ref{table:claim_averages}.

The reason for excluding the data of later runs is brevity and ease of review. The legitimacy of doing so is that from the $3rd$ run of the \texttt{demand} function, the gas costs stabilise and the remaining values are pretty much the same. In case of the \texttt{claim} function, the costs stabilise from the $2nd$ run. The rest of the data may be reviewed in the repository \cite{metin2020faucet}.

Added costs in the first $2$ of the \texttt{demand} function calls originate from the initialisation of the cyclic buffers. The number is $1$ for the \texttt{claim} function due to the fact that what it initialises is the balance vector, which is one dimensional, unlike the cyclic buffers.

These extra costs may totally be avoided showing up in the initial rounds of public functions, by initialising them in a dedicated epoch, in the launching of the system; e.g. in the present setting, a practical way may be to explicitly initialise them with placeholder values at the time of user registrations, that are carried out in the first epoch, before the first run of \texttt{demand} calls.

Thus, in essence, the representative values for the \texttt{demand} and \texttt{claim} functions are the ones that begin with the $3rd$ and $2nd$ runs of them, respectively. For this reason, in the regression analyses we referred to the $3rd$ run values of the functions. Nevertheless, we present the initial call values of these functions in the tables as they are collected in the tests.

Especially in the case of ordinary \texttt{claim} calls, i.e. those that do not trigger an epoch update, it is possible to have a near-perfect line fitting. Although the function includes control structures for checking the validity of the claim, the measured variability among successful ordinary claim calls is negligible under the experimental conditions.

A representative linear regression on the $3$rd run of the ordinary \texttt{claim} calls gives the equation:

\[
gas_c = 15,129.54 \cdot m + 38,986.67
\]

\noindent with a goodness of fit measure of $R^2 > .999999$, where $m$ is the number of resources, and $gas_c$ represents the resulting gas cost of an ordinary \texttt{claim} call.

For the \texttt{demand} function, it is still possible to have a near-perfect line fitting, with negligibly small deviations. This is due to the branching in the decision on the $ds$ values, since the position at which it will appear, thus the number of updates, vary among the different demand vectors. For example, the best case is the appearance of the $ds$ in the first component of the demand vector, since it will not be updated, and always the main branch will be taken. The worst case for this comparison is met when the demands are in strictly ascending order of fractional demand values and their scaled reciprocals are strictly decreasing, in which case an update is needed at each subsequent step.

\begin{table}[h!]
	\resizebox{\columnwidth}{!}{
		\begin{tabular}{|c|c|c|c|c|}
			\hline
			& \multicolumn{4}{c|}{Demand Averages} \\
			\hline
			$m$ & 1 & 2 & 3 & St.Dev. Avg. \\
			\hline
			2 & 140,675.700 & 141,179.800 & 74,353.900 & 7,380.222 \\
			\hline
			5 & 232,787.700 & 233,565.000 & 114,180.000 & 10,174.391 \\
			\hline
			10 & 386,698.000 & 385,129.300 & 185,793.500 & 16,013.621 \\
			\hline
			20 & 690,374.000 & 690,633.100 & 325,387.800 & 26,510.636 \\
			\hline
			30 & 994,028.000 & 993,250.700 & 453,904.900 & 36,522.665 \\
			\hline
			40 & 1,298,762.400 & 1,298,503.300 & 594,627.100 & 46,788.951 \\
			\hline
			50 & 1,601,121.900 & 1,601,381.000 & 727,436.600 & 58,975.615 \\
			\hline
			100 & 3,119,320.800 & 3,121,393.600 & 1,398,299.700 & 112,798.587 \\
			\hline
		\end{tabular}
	}
	\caption{Mean gas costs of the first three rounds of demand calls, and mean within-round standard deviation over all ten rounds. All configurations have $10$ users and a replenishment of $1,500$ units per resource per epoch.}
	\label{table:demand_averages}
\end{table}

\begin{figure}[!htbp]
	\includegraphics[width=\columnwidth]{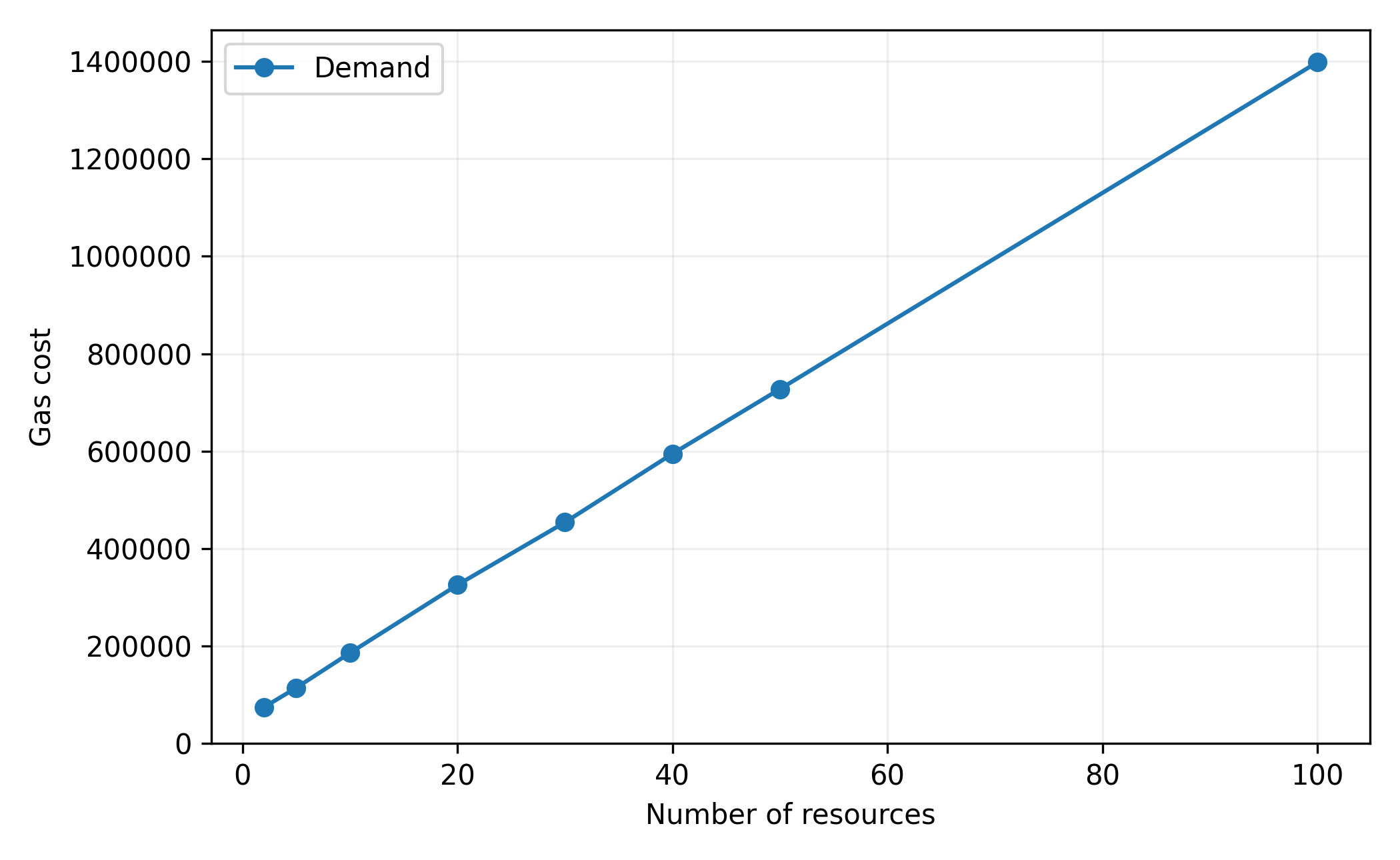}
	\caption{Demand function for varying number of resources (third run)}
	\label{demand_3_line}
\end{figure}

Still the variance is very low, as can be seen in the last column of Table \ref{table:demand_averages}. The values in this column are the averages of standard deviations from all $10$ runs of the tests.

A representative linear regression on the $3rd$ run of the \texttt{demand} function gives the equation:

\[
gas_d = 13,507.40 \cdot m + 50,322.81
\]

\noindent with a goodness of fit measure of $R^2 = .999943$, where $m$ is the number of resources, and $gas_d$ represents the resulting gas cost of the \texttt{demand} function.

\begin{figure}[!htbp]
	\includegraphics[width=\columnwidth]{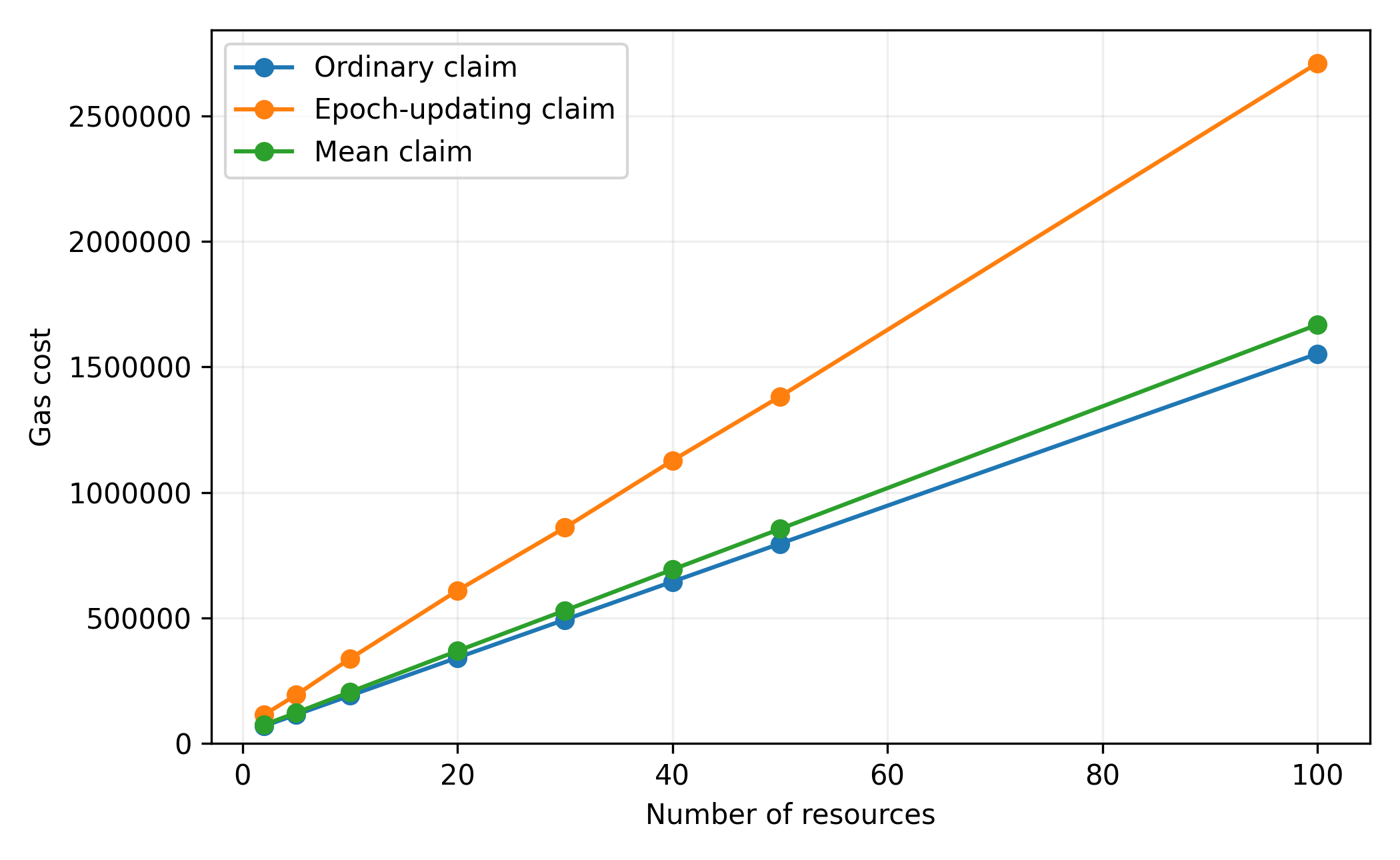}
	\caption{Gas costs of claim calls with and without epoch updates for varying number of resources (third run)}
	\label{claim_3_line}
\end{figure}

\begin{table}[h!]
	\resizebox{\columnwidth}{!}{
		\begin{tabular}{|c|c|c|c|c|}
			\hline
			& \multicolumn{4}{c|}{Ordinary Claim Averages} \\
			\hline
			$m$ & 1 & 2 & 3  & St.Dev. Avg. \\
			\hline
			2 & 114,166.000 & 69,166.000 & 69,166.000 & 0.000 \\
			\hline
			5 & 204,644.000 & 114,644.000 & 114,644.000 & 0.000 \\
			\hline
			10 & 355,294.000 & 190,294.000 & 190,294.000 & 0.000 \\
			\hline
			20 & 656,594.000 & 341,594.000 & 341,594.000 & 0.000 \\
			\hline
			30 & 957,894.000 & 492,894.000 & 492,894.000 & 0.000 \\
			\hline
			40 & 1,259,194.000 & 644,194.000 & 644,194.000 & 0.000 \\
			\hline
			50 & 1,558,827.333 & 795,494.000 & 795,494.000 & 500.000 \\
			\hline
			100 & 3,066,906.000 & 1,551,906.000 & 1,551,906.000 & 0.000 \\
			\hline
		\end{tabular}
	}
	\caption{Mean gas costs of the first three rounds of ordinary claim calls, and mean within-round standard deviation over all ten rounds. All configurations have $10$ users and a replenishment of $1,500$ units per resource per epoch.}
	\label{table:claim_averages}
\end{table}

The second terms of the regressions, $\beta_2$, represent the estimated constant costs in the execution of the functions, and the coefficients of the first terms, $\beta_1$, represent the estimated marginal costs per resource type. Although both functions perform three resource-wise operations, the marginal cost of the \texttt{claim} function is slightly higher than that of the \texttt{demand} function\footnote{The resource-wise operations are represented in lines $16$, $17$, $20$ ($24$) of Algorithm \ref{demand_pseudocode}, and lines $22$-$24$ of Algorithm \ref{claim_pseudocode}, respectively. In Solidity, the three claim operations are combined in a single loop, whereas demand uses three separate loops. The relative complexity of the share calculation in line $22$ may contribute to the difference between the coefficients.}.

Since the \texttt{update state} function is executed within the public functions, its gas cost is estimated by subtracting the average gas cost of ordinary \texttt{claim} calls from that of the \texttt{claim} call triggering the epoch update in the same round.

At the beginning of each public function execution \texttt{update state} is called, but only in the condition of epoch update does the function execute fully. In other cases, the function merely checks the epoch number against the block number, and discovering there is no need for update, it returns. The cost of checking epoch state and returning when no epoch update is needed, can be considered part of the public functions' execution.

A representative linear regression on the $3$rd run gives the equation:

\[
gas_{us} = 11,283.58 \cdot m + 28,792.38
\]

\noindent with a goodness of fit measure of $R^2 = .999635$, where $gas_{us}$ represents the incremental gas cost of an epoch update.

The two resource-wise operations of \texttt{update state} are shown in lines $14$ and $18$ of Algorithm \ref{udate_state_pseudocode}. Its incremental gas cost has the lowest coefficient, $\beta_1$.

\begin{figure}[!htbp]
	\includegraphics[width=\columnwidth]{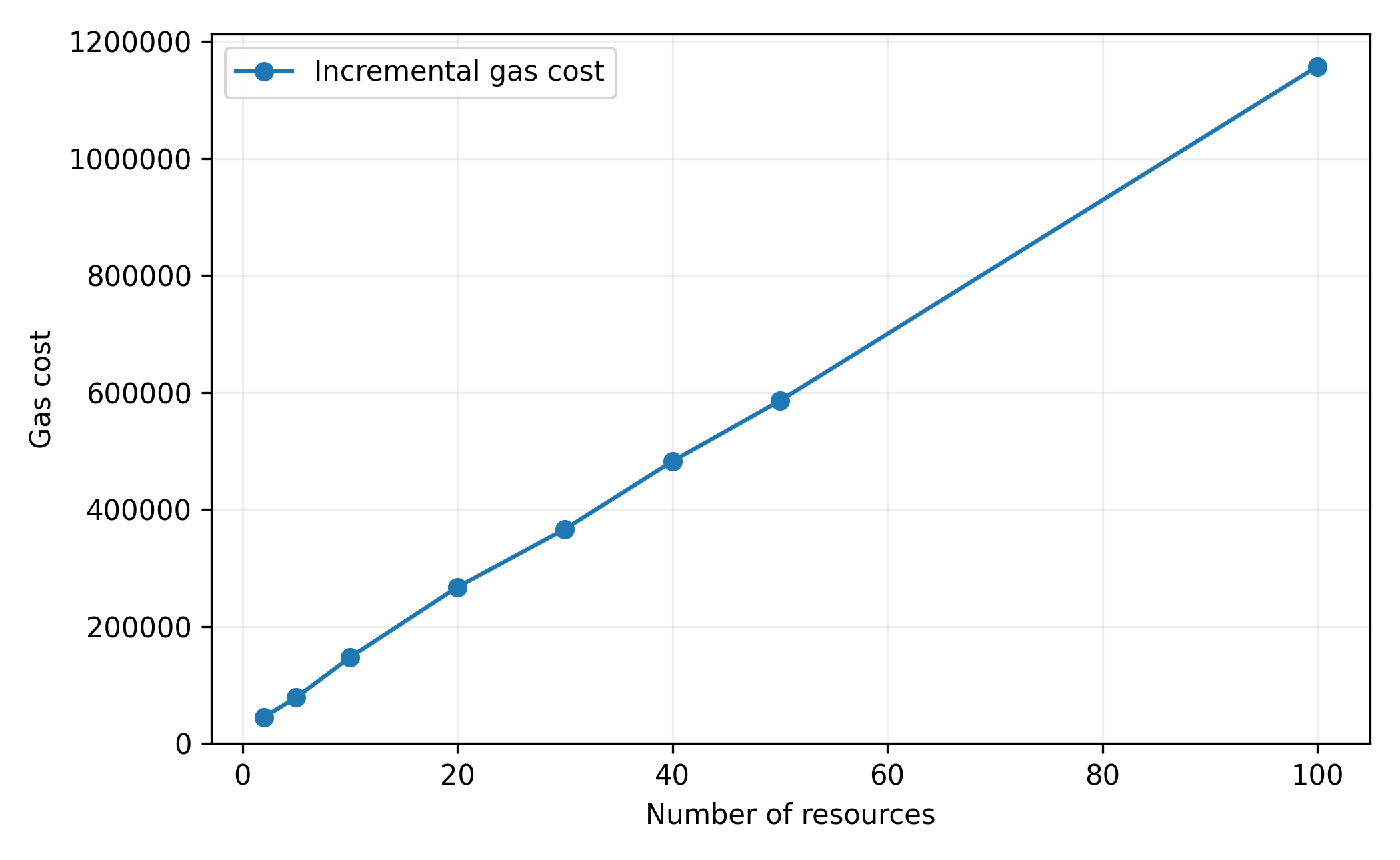}
	\caption{Incremental gas cost of epoch updates for varying number of resources (third run)}
	\label{update_state_3_line}
\end{figure}

\begin{table}[h!]
	\resizebox{\columnwidth}{!}{
		\begin{tabular}{|c|c|c|c|c|}
			\hline
			& \multicolumn{4}{c|}{Epoch Update Overhead} \\
			\hline
			$m$ & 1 & 2 & 3 & Average (10) \\
			\hline
			2 & 60,037.000 & 40,811.000 & 45,037.000 & 45,269.200 \\
			\hline
			5 & 89,339.000 & 78,565.000 & 78,565.000 & 80,487.600 \\
			\hline
			10 & 149,455.000 & 138,681.000 & 147,133.000 & 141,448.800 \\
			\hline
			20 & 273,913.000 & 263,139.000 & 267,365.000 & 255,764.400 \\
			\hline
			30 & 381,467.000 & 366,467.000 & 366,467.000 & 362,050.600 \\
			\hline
			40 & 510,151.000 & 478,247.000 & 482,473.000 & 478,056.600 \\
			\hline
			50 & 610,919.667 & 605,027.000 & 585,801.000 & 594,461.467 \\
			\hline
			100 & 1,163,927.000 & 1,157,379.000 & 1,157,379.000 & 1,155,498.200 \\
			\hline
		\end{tabular}
	}
	\caption{Incremental gas cost of epoch-updating claim calls relative to the mean ordinary claim cost in the same round, for the first three rounds and averaged over all ten rounds. All configurations have $10$ users and a replenishment of $1,500$ units per resource per epoch.}
	\label{table:update_state_averages}
\end{table}

Initialising storage arrays at deployment time by the \texttt{constructor} function is yet another way of avoiding added costs in the initial executions of the functions. The problem with this approach is that it can lead to increased \texttt{constructor} function gas cost numbers. The point is, unlike the other loops investigated in the present study, storage array initiation is a fixed cost, expended once for the system, and it is well distributable over transactions. Therefore it is not a bottleneck, neither is it a part of the algorithm complexity. Our particular design is one among the many possible, preferred for its convenience.

\begin{figure}[!htbp]
	\includegraphics[width=\columnwidth]{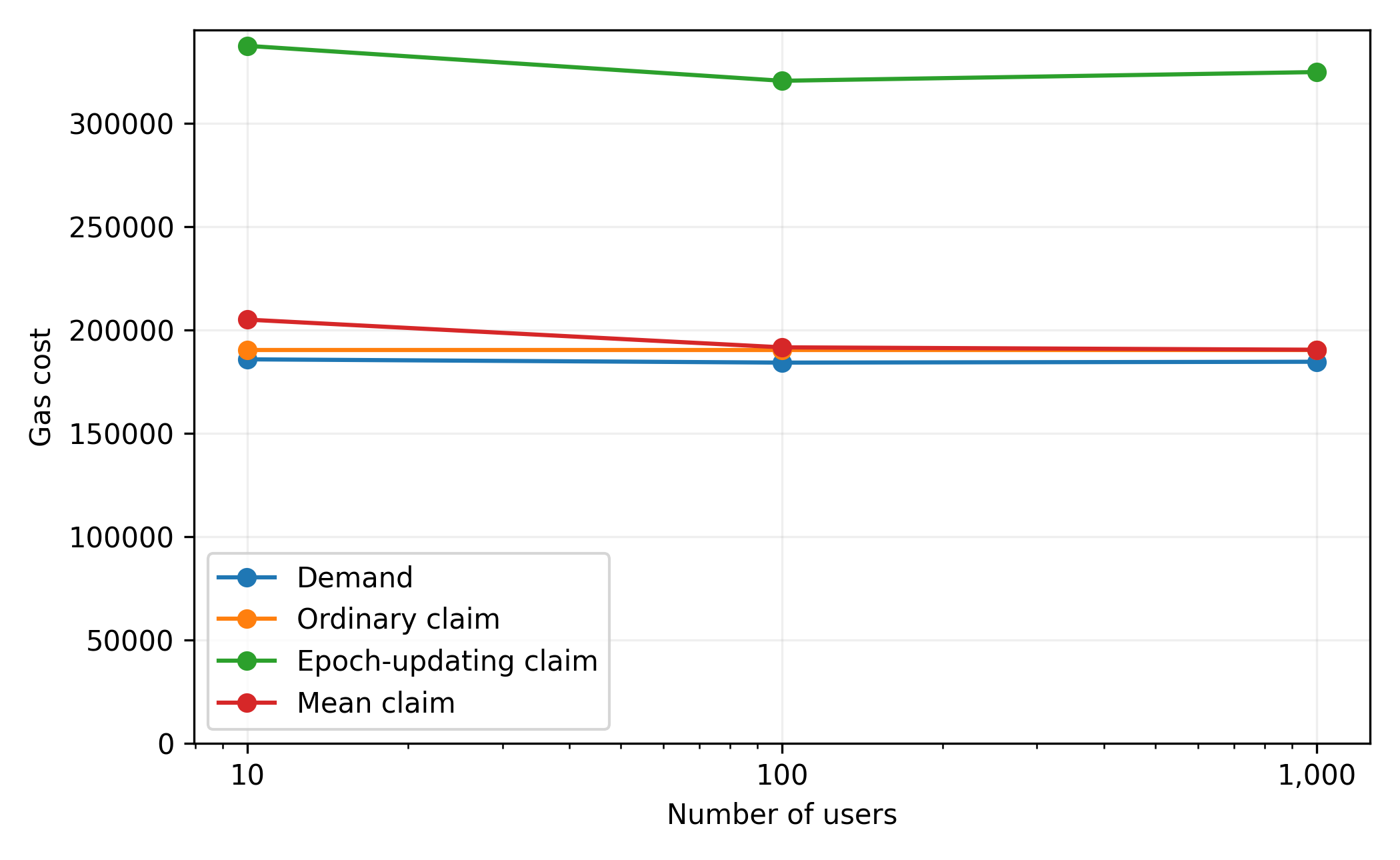}
	\caption{Gas costs for varying number of users (third run)}
	\label{user_scaling}
\end{figure}

In contrast to the number of resources, the number of users does not appear as a factor in the computational complexity of the functions. As shown in Figure \ref{user_scaling}, increasing the number of users from $10$ to $1,000$, while keeping the number of resources fixed at $10$, has no systematic effect on the gas costs of individual function calls. In particular, the gas cost of ordinary \texttt{claim} calls remains exactly the same, at $190,294$ gas, in all three configurations in the representative round.

A representative linear regression on the gas costs of \texttt{claim} calls triggering an epoch update gives the equation:

\[
gas_{c+us} = 26,413.13 \cdot m + 67,779.06
\]

\noindent with a goodness of fit measure of $R^2 = .999933$. Considering a hypothetical block gas limit of $32,000,000$, rather than the experimental limit, extrapolation of this regression suggests that ADRF can run with more than $1,000$ resource types, even when an epoch update and a claim are executed within the same transaction.

Representative gas cost measurements for the \texttt{demand} and \texttt{claim} functions and the incremental cost of epoch updates can be seen in Figures \ref{demand_3_line}, \ref{claim_3_line} and \ref{update_state_3_line}, respectively.

\section{Discussion}
\label{discussion}

The results and regression-based extrapolation suggest that ADRF can be efficiently run on blockchains with hundreds of resource types, without the gas costs of individual function calls growing with the number of users. For computer resources, this number may seem redundant, but the resource allocation problems that blockchain systems can address enjoy a wider scope. Also in computer systems context, the number may grow over the non-physical resources such as shared variables and locks.

One particular example is the token economies that are common in blockchains. Fungible and non-fungible tokens are distributed in many use cases, and questions of fair distribution are faced often. A system like ADRF may offer opportunities of securing the fairness of distribution among different types of tokens with different reserves, in such cases.

We also would like to argue in this context that DRF, and in turn ADRF, is an implicit pricing mechanism, by naturally valuing most demanded and least supplied resources over the others, since higher demand volumes and lower resource reserves increase the likelihood of a given resource determining the dominant share of a user.

\section{Conclusion}
\label{conclusion}

In the present study, we adapted Precomputed Dominant Resource Fairness to the blockchain context. The algorithm approximates the Dominant Resource Fairness allocation through PDRF, with measured gas costs growing linearly with the number of resource types and independently of the number of users. As such, it can be used as a solution to problems including fair allocation of different resources among users with heterogeneous needs, within the blockchain ecosystems.

\section*{Declaration of AI Assistance}

In the present version of this manuscript, OpenAI's ChatGPT was used to assist in the revision of the study in November 2026. This collaboration included reviewing the mathematical formulation and Solidity implementation, identifying implementation errors, replicating and extending the computational experiments, re-analysing and re-visualising the updated experimental results, and reviewing and typesetting the manuscript. The author retains responsibility for the scientific content and conclusions of the study.

\bibliographystyle{elsarticle-num} 
\bibliography{bibl}





\end{document}